\documentclass{epl}
\title{The unbinding transition of mixed fluid membranes}
\shorttitle{The unbinding transition etc.}
\author{Shigeyuki Komura\inst{1}\thanks{E-mail: 
\email{komura@comp.metro-u.ac.jp}}
        \and
        David Andelman\inst{2}\thanks{E-mail: 
\email{andelman@post.tau.ac.il}}}
\institute{
  \inst{1} Department of Chemistry, Faculty of Science,
           Tokyo Metropolitan University,
           Tokyo 192-0397, Japan \\
  \inst{2} School of Physics and Astronomy,
           Raymond and Beverly Sackler Faculty of Exact Sciences,
           Tel Aviv University, Ramat Aviv 69978, Tel Aviv, Israel
}

\begin{document}

\maketitle
\begin{abstract}
A phenomenological model for the unbinding transition
of multi-component fluid membranes is proposed,
where the unbinding transition is described using a
theory analogous to Flory-Huggins theory for polymers.
The coupling  between the lateral phase
separation of inclusion molecules and the membrane-substrate distance
explains the phase coexistence between two unbound phases
as observed in recent experiments by Marx \etal \ [Phys.\
Rev.\ Lett.\ {\bf88}, 138102 (2002)].
Bellow a critical end-point temperature, we find that the unbinding
transition becomes first-order for multi-component membranes.
\end{abstract}

\section{Introduction}

Adhesion of membranes and vesicles is responsible for cell-cell
adhesion which plays an important role in all multicellular organisms.
In general, bio-adhesion is governed by the interplay of
a variety of generic and specific interactions~\cite{LS}.
The latter interaction acts between complementary pairs of
proteins such as ligand and receptor, or antibody and antigen.
Well-studied examples of such coupled systems are
biotin-avidin complex~\cite{LISK}, or selectins and their
sugar ligands~\cite{ZB95}.

The problem of adhesion of multi-component membranes is intimately
related to that of domain formation.
Experimentally, adhesion-induced lateral phase separation has been
observed for various systems~\cite{AFS,NFBS,BBS},
and it was reported that adhesion molecules
aggregate spontaneously and form tight adhering domains. From the 
theoretical point of view, this problem has been
considered in~\cite{KA} using a phenomenological model,
where the inter-membrane distance is coupled to the
concentration of sticker molecules on the two adhering
membranes.
In a different approach, a lattice model for  a multi-component
membrane in contact with another substrate was
proposed~\cite{Lipowsky}, and was extended using detailed
Monte Carlo simulations~\cite{WNL,WL}.
More recently, a work combining these two approaches has been
published~\cite{WAKL}.

In a recent experiment by Marx \etal~\cite{MSSB},
the role of a long-range repulsion due to thermal fluctuations
(Helfrich repulsion) of the adhering membranes has been addressed.
By using  Reflection Interference Contrast Microscopy~\cite{AFS}, 
a multi-component bilayer
membrane with added lipopolymers (modified DOPE lipid
with polyethylenoxide) and cholesterol is examined in the vicinity of
an attractive substrate.
Analyzing the probability distribution of the membrane-substrate
spacing for various multi-component membranes,  a phase
separation between distinct lipopolymer-poor and lipopolymer-rich
states was found, where both states are {\it unbound} from the substrate.
Although this phenomena is dynamic in nature, the authors regard it
as a lateral phase separation induced by the Helfrich repulsion.
However a clear physical description for the appearance of such
a phase coexistence has not yet been given.
Moreover, a {\it first-order} unbinding transition scenario is required
to account for the multiple time scales in the time series of
multi-component membrane fluctuations~\cite{MSSB}.

In this Letter, we propose a simple phenomenological model for
multi-component (mixed) fluid membranes which can undergo
{simultaneously} a lateral phase separation and an unbinding transition.
The {model relies} on the coupling  between  inclusion
concentration and membrane-substrate spacing.
The lateral phase separation of inclusions affects the second
virial coefficient of the unbinding transition, and is taken
into account in analogy to Flory-Huggins theory for 
polymers~\cite{deGennes}.
The proposed model is successful in describing the phase separation
between two unbound states as mentioned above.
Furthermore, depending on the interaction strength or the temperature,
the model  exhibits  phase coexistence between two bound
phases, or between a bound and an unbound phase.
The latter case provides support to the first-order nature of the unbinding
transition, as is anticipated in experiments~\cite{MSSB}.

\section{Unbinding transition}

\begin{figure}
\onefigure{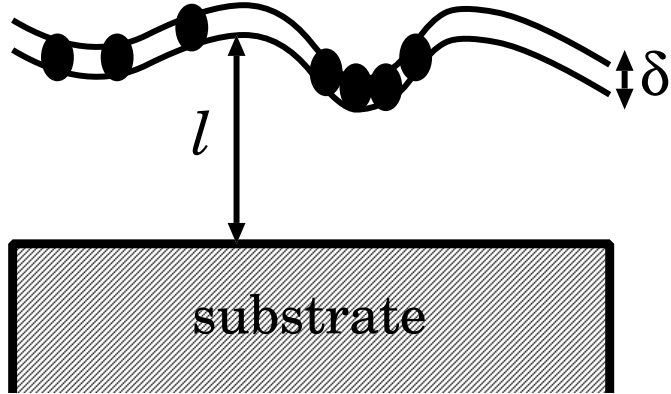}
\caption{A mixed fluid membrane adhering to a substrate.
Black filled ovals indicate inclusions such as proteins or
lipopolymers.
The height of the lower membrane leaflet from the
substrate is denoted by $\ell$.
The membrane thickness is $\delta$.}
\label{substrate}
\end{figure}
Fluid membranes in a lamellar stack or close to a substrate
experience steric repulsion arising from their reduced
undulation entropy due to the confinement effect.
Helfrich showed that the corresponding
interaction energy per unit area is given by~\cite{Helfrich78}
\begin{equation}
\label{undulation}
\label{steric}
v_{\rm s}(\ell) = \frac{b (k_{\rm B}T)^2}
{\kappa \ell^2},
\end{equation}
where $k_{\rm B}$ is the Boltzmann constant,
$T$ the temperature,
$\kappa$ the bending rigidity of the membrane
(of thickness $\delta$), and
$\ell$ the average height of the lower membrane lipid leaflet
from the substrate (see fig.~\ref{substrate}).
The numerical prefactor $b$ was calculated as
$b=3\pi^2/128$ in~\cite{Helfrich78} but its value is
still debatable in the literature~\cite{SO}.
The combination of the above steric repulsion and other
direct microscopic (van der Waals, electric, hydration)
interactions determines whether the membrane binds to the
substrate or unbinds.
Using functional renormalization-group techniques, Lipowsky and
Leibler showed that this unbinding transition is
{\it second-order}. 
The average spacing $\ell$
diverges as the strength of the attractive van der
Waals interaction $W$ (known as the Hamaker constant)
approaches a critical value from
above, i.e., $\ell \sim (W-W_{\rm c})^{-\psi}$ with
$\psi \approx 1.00$, where $W_{\rm c}$ is the critical strength of
Hamaker constant~\cite{LL1,LL2}.
It should be emphasized, however, that a simple superposition
of the Helfrich repulsion, eq.~(\ref{steric}), and other
direct interactions gives an incorrect (first-order)
description of this unbinding transition.
This failure is linked to the fact that eq.~(\ref{undulation})
was originally derived under the assumption that membranes
do not interact through any other forces.

Subsequently, a much simpler theory for the unbinding transition
was considered by Milner and Roux~\cite{MR}, and it
is briefly reviewed here.
Following the spirit of Flory-Huggins theory for
polymers~\cite{deGennes},
the Helfrich estimate of the entropy is taken into
account accurately, whereas the other interactions are
approximately incorporated via a second virial term.
The resulting free energy per unit area of the membrane
is~\cite{area}
\begin{equation}
\label{milnerroux}
f(w) = - k_{\rm B}T \chi\delta  w^2 +
\frac{b (k_{\rm B}T)^2}{\kappa \delta^2} w^3,
\end{equation}
where $w = \delta/\ell\ge 0$ cannot be negative.
In the above, $\chi$ is a second virial coefficient describing
the correction to the hard-wall repulsive interaction represented by the
cubic term in $w$.
This approach is justified since the perturbative effect of the 
direct interactions is included through a virial expansion rather
than the simple superposition of interactions.

Consider a membrane patch of volume $\nu = a^2 \delta$,
where $a = \sqrt{\kappa/k_{\rm B}T} \delta$ is the in-plane
cutoff.
When the enthalpic interactions between the membrane patch and
the substrate is small, $\chi$ is given by
\begin{equation}
\label{virial}
\chi = - \frac{1}{2 \nu^2}
\int {\rm d}^3\mbox{\boldmath $r$} \,
(1-\exp[-V_{\nu}(\mbox{\boldmath $r$})/k_{\rm B}T]),
\end{equation}
where $V_{\nu}(\mbox{\boldmath $r$})$ is the direct interaction energy,
and the integral is limited to positions such
that the membrane patches do not overlap.
The free energy (\ref{milnerroux}) has a continuous second-order
transition at $\chi=0$ between a bound state ($w > 0$) for
$\chi>0$ and an unbound state ($w = 0$) for $\chi < 0$.
Because Milner and Roux model predicted that $\chi \sim (W-W_{\rm c})$,
eq.~(\ref{milnerroux}) is also successful in describing
the critical exponent $\psi = 1$~\cite{MR}.

\section{Unbinding transitions of mixed membranes}

So far, the unbinding transition of a single-component
membrane has been discussed.
In the case of multi-component membranes,
the lateral phase separation affects the direct interactions
between the membrane and the substrate, and hence their
unbinding behavior.
For  simplicity, we consider a two-component membrane
adhering to a substrate as in fig.~\ref{substrate}.
The overall membrane state is characterized by
the membrane average distance $\ell$
from the substrate.
The internal degree of freedom, on the other hand, corresponds
to the membrane composition.
In addition to the lipid component that is the main building block
of the membrane, we introduce a second component or an
``inclusion'' representing additional
lipids, proteins, cholesterol or lipopolymers also residing
on the membrane.
Let the concentration of these inclusions be denoted by
$\Phi$ ($0 \le \Phi \le 1$).
Here we discuss the case in which the interaction between
two inclusions is attractive leading to a condensation 
transition. 
Then, below a certain temperature, the mixed membrane
undergoes a first-order transition, where an inclusion-poor
phase coexists with an inclusion-rich phase as observed 
experimentally~\cite{KPHM}. 
The first-order transition terminates at a critical point,
having a critical concentration $\Phi_{\rm c}$.
The concentration difference is defined as
$\phi = \Phi - \Phi_{\rm c}$.

Using the ``Flory-Huggins theory'' of the unbinding transition
mentioned above, and taking into account the lateral
phase separation of inclusions close to the critical point,
we propose the following free energy expression of a mixed
membrane undergoing an unbinding transition:
\begin{equation}
\label{model}
f(\phi,w) = -\mu \phi + \frac{1}{2} t \phi^2 + \frac{1}{4} \phi^4
- \chi w^2 + \frac{1}{2} w^3
+ \phi w^2,
\end{equation}
with the constraint $w \ge 0$.
All energy terms have been scaled by
$2b(k_{\rm B}T)^2/\kappa \delta^2$, and have units of energy per unit area.
The first three terms in eq.~(\ref{model})
depend only on $\phi$ and correspond to the Landau free
energy of a two-component membrane undergoing a lateral
inclusion-lipid phase separation; $\mu$ is the chemical potential and
$t$ the reduced temperature.
The quartic coefficient has been chosen to be positive and
is arbitrary set to $\frac{1}{4}$ without loss of generality.
The next two terms depend only on $w$, and represent the
unbinding transition of the membrane (see eq.~(\ref{milnerroux})).
The last term is the lowest order coupling term between
$\phi$ and $w$ allowed by symmetry with a coefficient that
has been set to unity without loss of generality.
The physical meaning of this term is as follows.
When the mixed membrane is quenched into a two-phase region,
an inclusion-poor phase ($\phi < 0$) coexists with an
inclusion-rich phase ($\phi > 0$).
This can lead to different direct interactions $V_{\nu}$ and
hence different second virial coefficients $\chi$ for each of
the domains.
We model this situation by considering an effective
second virial term $-\chi_{\rm eff} w^2= -(\chi -\phi) w^2$,
which leads to the proposed coupling term $\phi w^2$.
Notice that this term being linear in $\phi$ changes its sign for
$\phi \rightarrow - \phi$.

In addition, a linear term in $w$,  arising from
an external force on the membrane such as pressure or gravity,
can also appear in eq.~(\ref{model}).
In the presence of such a term, the unbinding transition becomes
first-order even for a single-component membrane~\cite{MR}.
However, this situation will not be further considered here.
Another important assumption is that the phase separation of
the inclusion molecules does not affect the homogeneous bending
rigidity of the membrane $\kappa$.
In other words, the direct microscopic (van der Waals,
electrostatic, hydration) interactions are modified by the
phase separation, whereas the steric interaction remains unchanged.
Hence our model should be distinguished from the one in~\cite{NP},
where it was shown that the coupling between membrane curvature and
inclusion concentration leads to lateral phase separation.

Minimizing the free energy (\ref{model}) with respect
to $w$ gives $w=0$ (unbound) for $\chi< \phi $ and
$w=\frac{4}{3}(\chi - \phi) > 0$ (bound) for $\chi > \phi$.
By substituting back the value of $w$ into eq.~(\ref{model}),
the free energy as a function of $\phi$ only becomes
\begin{equation}
\label{freeenergy1}
f(\phi) = - \mu \phi +
\frac{1}{2} t \phi^2 + \frac{1}{4} \phi^4
~~~~~\mbox{for~~$\phi > \chi$}
\end{equation}
\begin{equation}
\label{freeenergy2}
f(\phi) =
-\frac{16}{27}\chi^3 + \left(\frac{16}{9} \chi^2 - \mu \right) \phi
+ \left( \frac{1}{2} t - \frac{16}{9} \chi \right) \phi^2
+\frac{16}{27} \phi^3 + \frac{1}{4}\phi^4
~~~~~\mbox{for~~$\phi < \chi$}
\end{equation}
depending on the value of $\chi$.
This free energy is continuous at $\phi = \chi$.
We next minimize this free energy with respect to $\phi$
to find the equation of state.
The two-phase region is obtained by the Maxwell construction,
and the whole phase diagram is calculated numerically.

\section{Phase behavior}

\begin{figure}
\twoimages[scale=0.4]{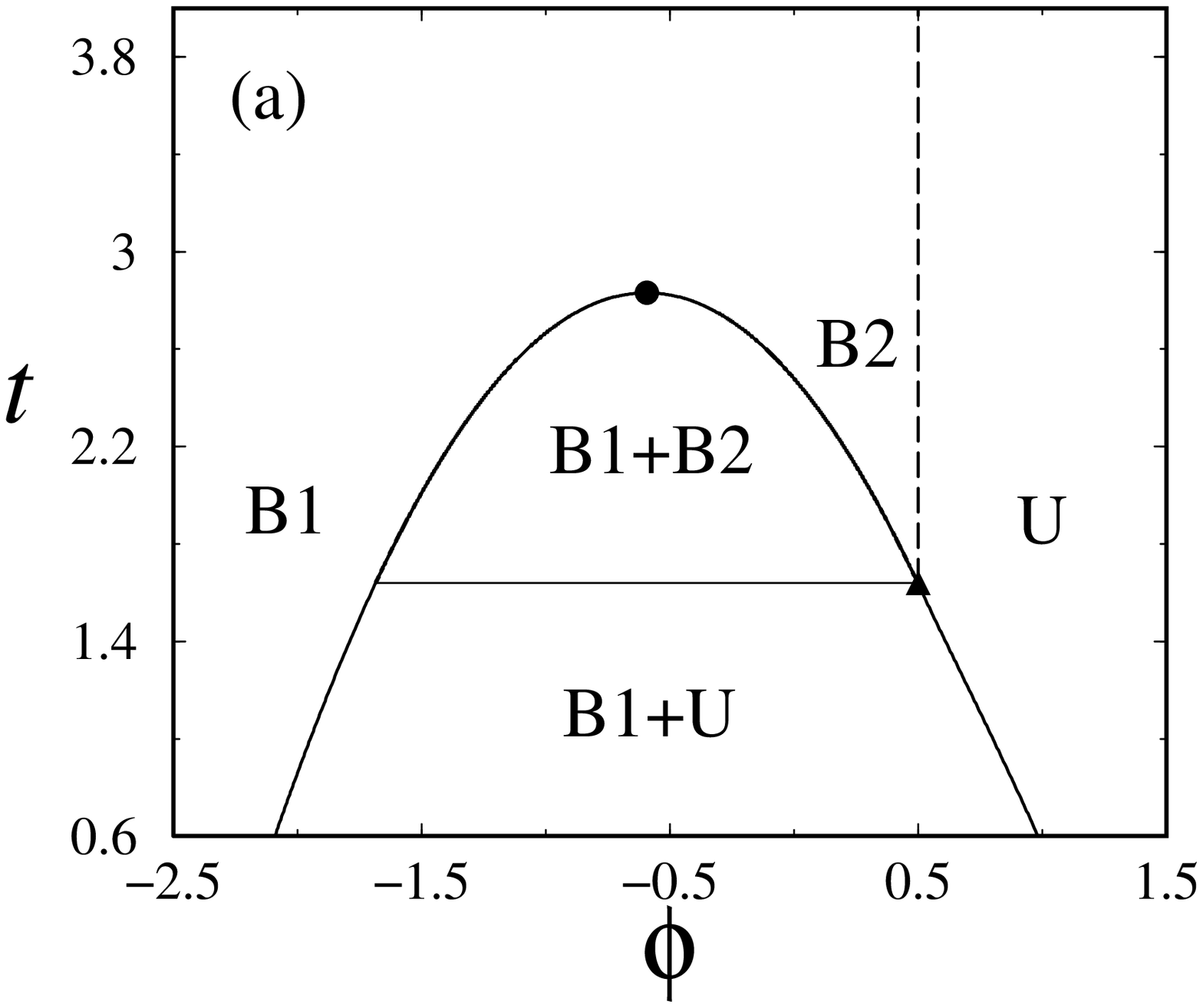}{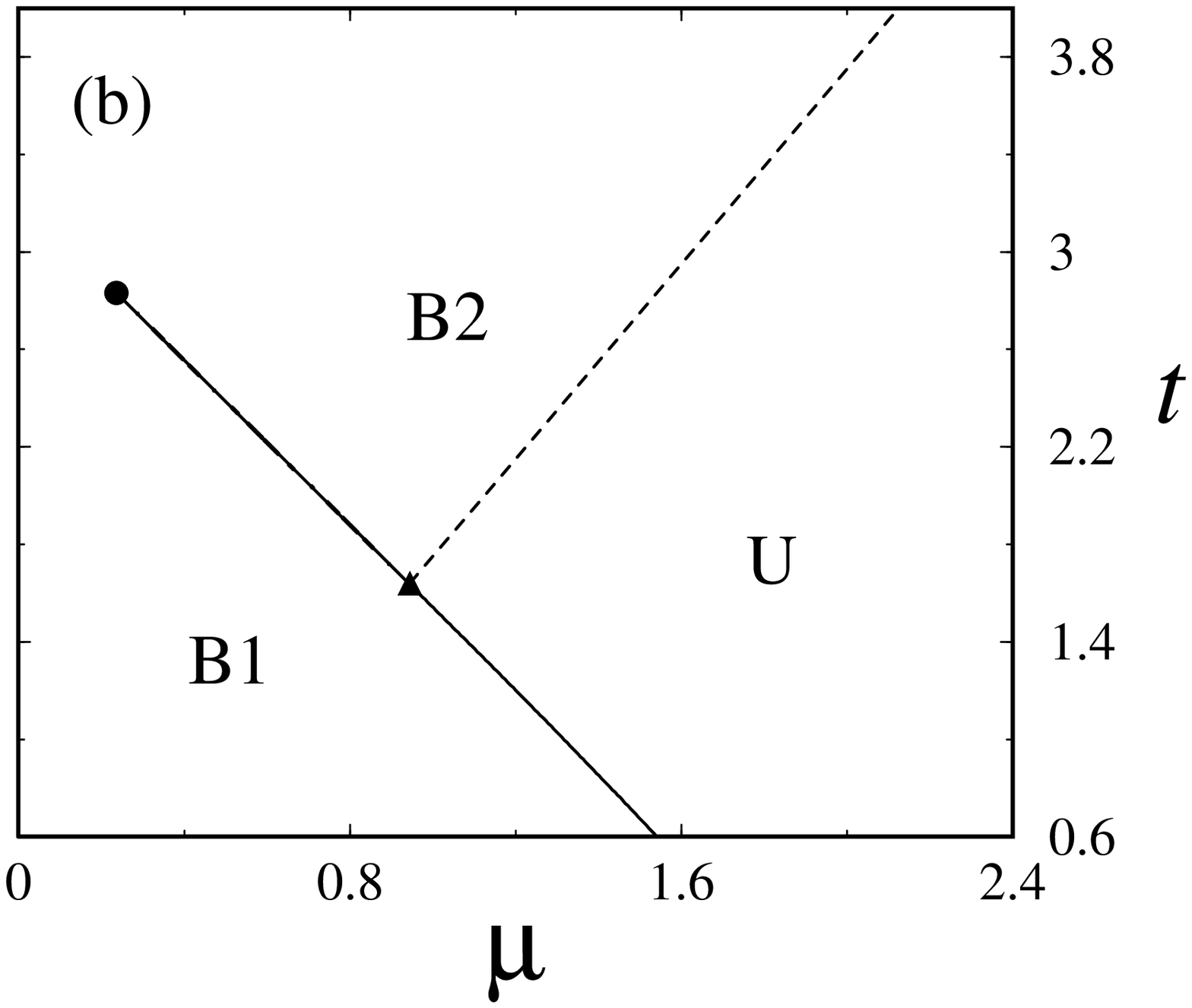}
\caption{The phase diagrams for $\chi = 0.5$ as a function
of (a) inclusion concentration $\phi$ and reduced temperature $t$,
and (b) inclusion chemical potential $\mu$ and reduced temperature
$t$.
The continuous line is a first-order line, whereas the
dashed line is a second-order one.
The critical point (CP) and the critical end-point (CEP) are indicated
by a filled circle and a triangle, respectively.
The bound ($w > 0$) and the unbound ($w = 0$) phases are denoted
as B and U, respectively.
Below the critical point, there is a coexistence region
either of the two bound phases (B1+B2), or a bound  and an
unbound phases (B1+U).
}
\label{phase05}
\end{figure}
The phase behavior of the above model depends on the value
of $\chi$.
For $\chi > -\frac{8}{27}$, a second-order transition line
(critical line) where $\phi=\chi$,
separates the unbound phase ($w=0$) from the
bound one ($w>0$), as shown in fig.~\ref{phase05}
for $\chi=0.5$.
The critical line ends at a critical end-point (CEP) on the
first-order transition line.
The first-order transition line itself ends at an ordinary critical
point (CP) corresponding to a liquid/vapor-type CP
between two bound phases (B1+B2).
Below the CEP temperature, the first-order line separates a bound
phase from an unbound phase (B1+U).

The critical point between the two bound phases is
given by $t_{\rm c} = \frac{32}{9}(\chi + \frac{8}{27})$ and
$ \phi_{\rm c} = -\frac{16}{27}$.
Hence, the critical temperature is increased from
$t_{\rm c}=0$ to $t_{\rm c} \approx 2.83$ for
$\chi=0.5$
due to the
coupling between  inclusion concentration $\phi$ and
membrane-substrate distance $\ell = \delta /w$.
In other words, the phase separation is {\it enhanced} by the
adhesion for $\chi > -\frac{8}{27}$.
This result is in accordance with the previous theoretical
models~\cite{KA,WNL,WL}.
Between the CEP and the CP of the two-phase region (B1+B2),
the two coexisting values of $\phi$ lead to different
membrane-substrate distances given by
$\ell = \frac{3}{4}\delta(\chi-\phi)^{-1}$.
Since the $\ell$ value for B1 is smaller than that for B2, the phases B1 and B2
correspond to ``tight'' and ``loose'' bound phases,
respectively, as discussed in~\cite{KA}.
We see that $\ell$ diverges as
$(\chi-\phi)^{-1}$ at $\phi=\chi$, and $w = \delta / \ell$
increases continuously from zero.

\begin{figure}
\twoimages[scale=0.4]{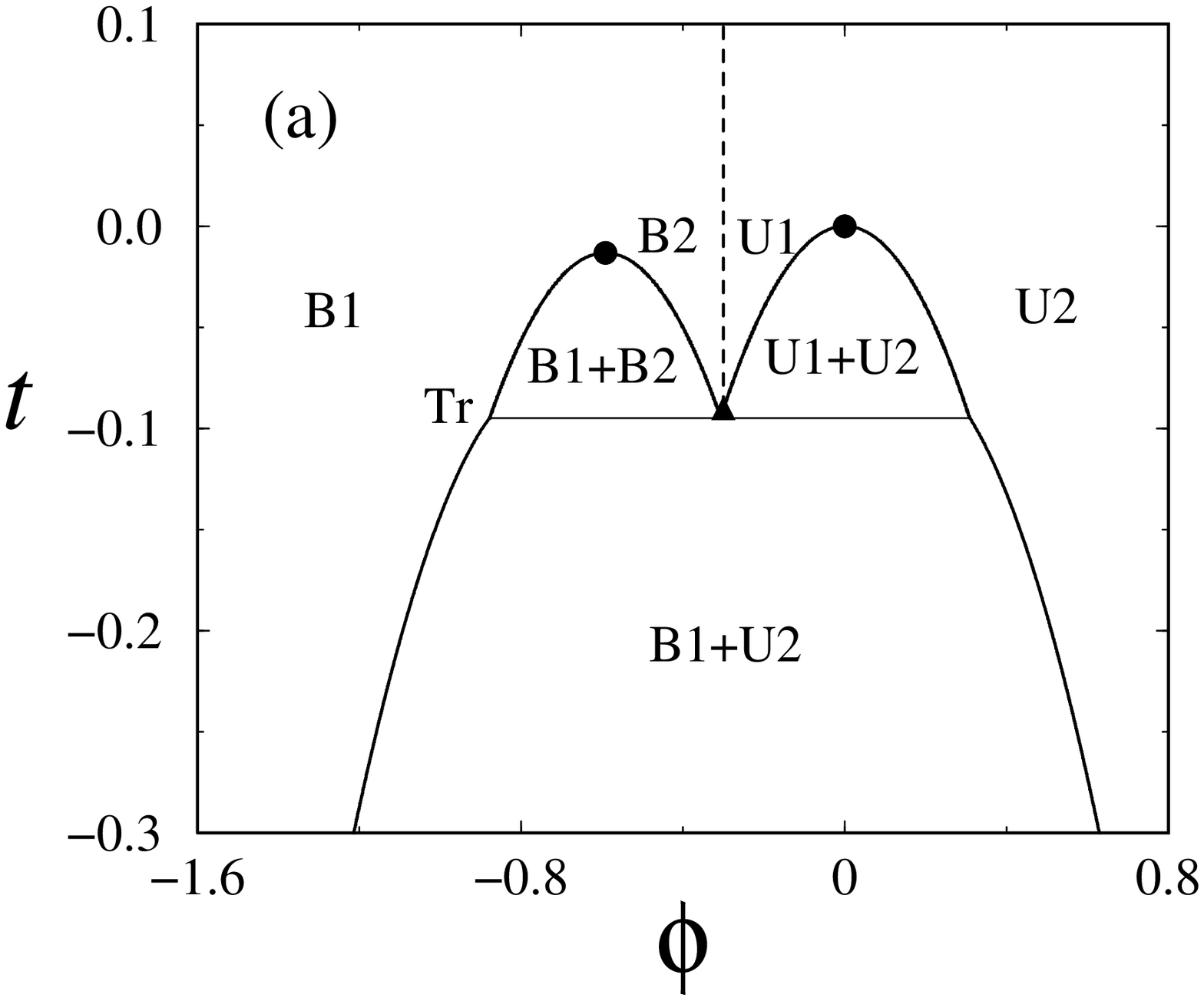}{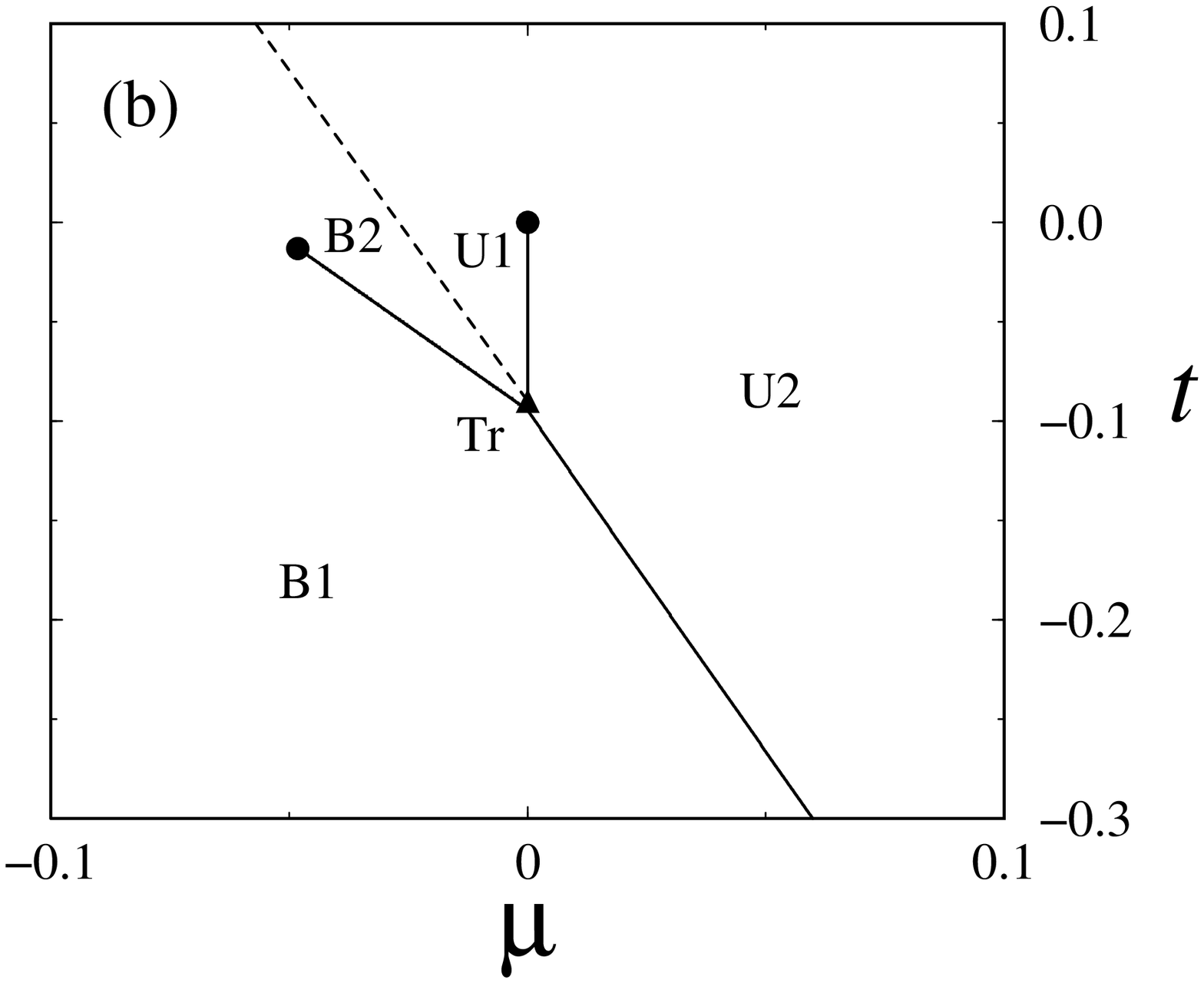}
\caption{The phase diagram for $\chi=-0.3$ as a function
of (a) inclusion concentration $\phi$ and reduced temperature $t$,
and (b) inclusion chemical potential $\mu$ and reduced temperature
$t$.
Same notation of different lines and symbols is used
as in fig.~\ref{phase05}.
In addition to the critical end-point (denoted by a filled
triangle), there are two critical points (denoted by filled
circles) and a triple point denoted by Tr.
In (b), the CEP and Tr are two distinct but very close points,
although it is hard to see it on the figure.
}
\label{phase-03}
\end{figure}
In fact, even the very simple free energy~(\ref{model}) leads
to a rich topology of the phase diagram.
In addition to the coexistence between two bound phases (B1+B2),
a new two-phase region between two unbound phases (U1+U2) appears
for $-\frac{32}{81} < \chi < -\frac{8}{27}$.
As an example, fig.~\ref{phase-03} gives the phase diagram for 
$\chi=-0.3$.
In this case, two different critical points are found:
one at $t_{\rm c} = \frac{32}{9}(\chi + \frac{8}{27})$, $\phi_{\rm c}
= -\frac{16}{27}$
as before, and the other at $t_{\rm c} = 0$, $\phi_{\rm c} = 0$.
Hence the phase separation is not necessarily enhanced.
There is also a triple point (Tr) at which two bound phases and
one unbound phase coexist (B1+B2+U2).
Below the triple point temperature, there is a region of
two-phase coexistence between a bound and an unbound phase
(B1+U2).

For $\chi < -\frac{32}{81}$, a similar $(\phi, t)$ phase diagram as
in fig.~\ref{phase05}(a) is obtained with the critical line
being located at $\phi < \phi_{\rm c}=0$ (not shown as a graph).
Below the critical point $t_{\rm c}=0$, $\phi_{\rm c} = 0$,
two unbound phases coexist until the CEP temperature.
Then a region of two-phase coexistence between a bound and
an unbound phase appears below the CEP.

As mentioned before, the coexistence between two unbound states
has been observed experimentally for mixed membranes with added
lipopolymers and cholesterol~\cite{MSSB}.
For the case of fixed cholesterol content,
distinct lipopolymer-poor  and lipopolymer-rich
states coexist, and both states are found in experiments to be
unbound.
Our model predicts that such a coexistence appears for
$\chi < -\frac{8}{27}$.
Note also that the unbinding transition becomes
first-order below the CEP or Tr, for any value of $\chi$.
This is in accordance with the experimental prediction~\cite{MSSB}.
However, the first-order and the second-order lines are connected
by a CEP and not by a tricritical point.

The present results can be compared with those obtained in other
theoretical works.
In our previous phenomenological model for the adhesion-induced
phase separation~\cite{KA}, we have considered only
the phase coexistence between two bound phases, and the unbinding
behavior of the membrane was not taken into account.
In that work we showed that the phase separation between two
bound phases is always enhanced by adhesion, whereas in the
present model, it can either be enhanced or suppressed, depending
on $\chi$.
Somewhat similar phase diagrams to  ours have been obtained in~\cite{WNL,WL}
by a  mean-field treatment of a lattice model.
In those works, the topology of the phase diagrams depends on several
quantities such as the sticker binding energy, the potential range,
or the strength of the {\it cis}-interaction between the stickers.
Some of their phase diagrams also include a two-phase
region where bound and unbound phases coexist.
However, the coexistence between two unbound phases has not been predicted
within the lattice model~\cite{WNL,WL}.

\section{Extensions of the model}

We finally discuss two possible extensions of our model.
The first situation deals with membranes having
finite tension $\sigma$.
Since the tension strongly suppresses the out-of-plane
fluctuations of the membranes, the steric interaction is weakened.
A self-consistent calculation of the Helfrich repulsion
under tension is given by~\cite{Seifert}
\begin{equation}
\label{udo}
v_{\rm s}(\ell;\sigma) =
\frac{b (k_{\rm B}T)^2}{\kappa \ell^2}
\left[
\frac{\ell/\lambda}{\sinh(\ell/\lambda)}
\right]^2,
\end{equation}
where $\lambda=(2 k_{\rm B} T/\pi \sigma)^{1/2}$.
Following the argument in~\cite{DC}, the generalized
free energy for the mixed membranes under tension
can be obtained by replacing the term $w^2/2$ in eq.~(\ref{model})
with $x^2 w^3/(2\sinh^2 x)$, where $x=\ell/\lambda$.
In general, tension induces binding of membranes, in cases where
membranes unbind
in the {\it absence} of tension.
Hence, for $x \gg 1$ (large tension), one expects that the region
of bound phases dominates most of the phase diagram.

Another interesting case is the adhesion between two fluctuating
mixed membranes.
Symmetry consideration with respect to the exchange of the two
membranes gives the allowed coupling terms between the inclusion
concentration on each of the membranes and the distance between them.
A rich phase behavior is expected when the average inclusion
concentration on the two membranes is not identical~\cite{KA}.

\section{Summary}

We have proposed a phenomenological model for
the unbinding transition of two-component fluid membranes.
The coupling between lateral phase separation and the
membrane-substrate distance provides a rich phase behavior.
Our model is successful in describing the first-order nature
of the unbinding transition in multi-component membranes,
as well as the phase separation between two unbound phases. It
may provide a theoretical explanation for the
experimental observations made by Marx \etal~\cite{MSSB}.

\acknowledgments

We thank S. Marx and E. Sackmann for useful discussions.



\begin{thebibliography}{0}
\bibitem{area}
In~\cite{MR}, the free energy per unit volume is given
since they consider a stack of membranes.
Here we multiplied the membrane thickness $\delta$ to
obtain the free energy per unit area of the membrane.
\end{thebibliography}
\end{document}